\documentclass[12pt]{article}
\usepackage{color,epsfig}
\usepackage{graphicx}
\definecolor{violet}{rgb}{0.4,0,0.2}
\definecolor{vert}{rgb}{0,0.6,0.0}
\definecolor{navy}{rgb}{0.0,0.0,0.6}
\definecolor{orange}{rgb}{0.6,0.4,0.0}
\definecolor{bleu}{rgb}{0.3,0.0,0.8}
\definecolor{spin}{rgb}{0.3,0.6,0.0}

\def\vv{{\color{violet}v}} 
 \def\phhi{{\color{violet}\varphi}}
\def\calK{{\color{violet}{\cal K}}}\def\sS{{\color{vert}S}}
\def\calN{{\color{violet}{\cal N}}}

\def\nn{{\color{vert}n}}
\def\kk{{\color{vert}k}}

\def\mm{{\color{red}m}}
\def\mmu{{\color{red}\mu}}\def\hba{{\color{red}\hbar}}
\def\sigm{{\color{red}\sigma}}

  \def\VV{{\color{red}V}}

\def\calE{{\color{red}{\cal E}}}

\def\alp{{\color{violet}\alpha}}

\def\calR{{\color{black}{\bf T }}}

\def\bfK{{\color{vert}{\bf K }}}

\def\calV{{\color{bleu}{\cal V }}}
\def\gl{{\color{vert}{\ell }}}
\def\Bl{{\color{black}{\ell }}}
\def\le{{\color{bleu}{e }}}
\def\lle{{\color{black}{ l }}}
\def\ita{{\color{bleu} {\it a} }}
\def\itb{{\color{bleu} {\it b} }}

\def\bb{\large\color{black}  $ }  \def\fb{ $  }
\def\be{\large\begin{equation}\color{black} }
\def\fe{\end{equation}}
\def\rf{\color{black} (\ref }
\def\fr{)\,\color{navy} }

\def\rmn{ {\rm n}} \def\rmp{{\color{orange}\rm p}}
 \def\rmb{ {\rm b}}

\def\spose#1{\hbox to 0pt{#1\hss}}\def\lta{\mathrel{\spose{\lower 3pt\hbox
{$\mathchar"218$}}\raise 2.0pt\hbox{$\mathchar"13C$}}}  \def\gta{\mathrel
{\spose{\lower 3pt\hbox{$\mathchar"218$}}\raise 2.0pt\hbox{$\mathchar"13E$}}}

\begin{document}

\title{\bf Band structure effects for dripped neutrons
in neutron star crust}

\author { Nicolas Chamel \footnote{email: Nicolas.Chamel@obspm.fr} \\ \hskip 1 cm\\   \\
  Laboratoire de L'Univers et ses TH\'eories (LUTH) \\ Observatoire de Paris, 92195 Meudon, France }

\maketitle

\color{navy}

\bigskip
{\bf Abstract.}
The outer layers of a neutron star are supposed to be formed of a solid
Coulomb lattice of neutron rich nuclei. At densities above
neutron drip density (about one thousandth of nuclear saturation density),
this lattice is immersed in a neutron fluid. Bragg scattering of those dripped neutrons by the nuclei
which has been usually neglected is investigated, within a simple mean field model with Bloch type
boundary conditions. The main purpose of this work is to provide some estimates for the
entrainment coefficients, as required for hydrodynamical two fluid simulations of neutron star crust \cite{CCHI, CCHII},
which relate the momentum of one fluid to the particle currents of the other two fluids \cite{AB76}. The implications
for the equilibrium neutron star crust structure are also briefly discussed.

\vskip  1 cm

\section{Introduction}
In a typical neutron star (with mass about one and a half that of the sun), a solid phase of neutron rich nuclei, hence forming the crust, is expected to occur
from the surface down to about {\bb \simeq 1\, \rm km\fb} in depth where the pressure and density are so high that the nuclei may be strongly deformed
and adopt ``exotic'' shapes \cite{PR95}. Inside the crust, above the neutron drip density {\bb \rho_{\rm drip}\simeq 10^{11}\, {\rm g}.{\rm cm}^{-3}\fb}, free neutrons are
found to coexist with the nuclei. Below the crust, the nuclei merge into a uniform mixture of nucleons and perhaps of some other mesons.

Bragg scattering of the dripped neutrons by the nuclei, which is analogous to the scattering of valence electrons by ions in ordinary solids, has been usually neglected
whereas it may be of quantitative importance for evaluating the processes involved in phenomena such as pulsar glitches. This issue has been pointed out by
a few authors \cite{Yak2001}\cite{Magierski2001}\cite{MMM2004} and band effects, namely the opening of band gaps in the single particle
energy spectrum, have been actually estimated in a simplified one dimensional case \cite{OyYam94}.

In a preceeding work \cite{CCHI, CCHII} we have introduced a macroscopic effective mass {\bb \mm_\star \fb} relevant for hydrodynamical simulations, such that the momentum of the neutron superfluid is given by the mean neutron velocity times this effective mass
in the crust rest frame. In a (super)fluid mixture, in general the momentum of one species is a linear combination of the particle currents of the other components. This is known as the entrainment
effect \cite{AB76}. We have shown how to obtain these entrainment coefficients in a two fluid neutron star model from the effective mass {\bb \mm_\star \fb}. Our analysis is complementary to
the two fluid description of neutron star core, based on the Fermi liquid theory (see Borumand et al \cite{BJK96} and references therein),
whose relativistic generalisation has been recently performed by Comer et al \cite{CJ2003} via a {\bb \sigma-\omega \fb} mean field model.

Particle transport in the inner crust is induced by displacements of the whole superfluid and it can thus be seen \cite{CCH2} that it is therefore not very sensitive to the superfluid energy gap unlike thermal properties
such as the specific heat. The main effect of the pairing interaction is actually to lead to a smearing of the
sharp neutron Fermi surface. Since at the densities of interest, the pairing gap is typically much smaller than the Fermi energy, corrections of the effective mass due to
pairing correlations are expected to be small. We shall therefore neglect as a first approximation pairing interaction (as in the works of Borumand et al \cite{BJK96} and Comer et al \cite{CJ2003}).

Since the pionnering Hartree-Fock calculations of Negele and Vautherin \cite{NV73}, more or less arbitrary boundary conditions have usually been applied. In particular the so called  ``Wigner-Seitz approximation'',
whereby nuclei are treated as if isolated,
has been widely used.  Whereas this approximation would be
appropriate for describing nucleons that are clustered around lattice nodes, it is more questionable
for the dripped neutrons which are delocalized over the crust. This means that one has necessarily to consider the whole medium and apply global boundary conditions. In the case of a periodic lattice of nuclei,
as a result of the well known Floquet-Bloch theorem \cite{Ashcroft81}, the problem can still be reduced to the so called Wigner-Seitz cell (not to be confused with the W-S sphere)
namely some ``elementary'' polyhedron surrounding a nuclei whose shape is completely determined by the lattice structure, supplemented with Bloch type boundary conditions. The single particle energy spectrum is composed of a set of analytic functions {\bb {\calE}_\alp\{\bf \kk \} \fb}
(in the following we shall use brackets for functional dependence in order to avoid confusions with mere multiplication)
 of the momentum {\bb \bf \kk \fb} (related to the translational symmetry), each sheet in momentum space being specified by a band index {\bb \alp \fb}
 (associated with rotational symmetry).

Recently Magierski et al \cite{Magierski2001} have suggested that shell effects arising from ``unbound'' neutrons
may be important for neutron star crust structure and have carried out Skyrme Hartree-Fock calculations
 in a cubic box with periodic boundary conditions. However such a choice of boundary conditions does not properly account for Bragg scattering since the ensuing single particle
 states are only those associated with vanishing Bloch momentum. Besides the Wigner-Seitz cell, which is taken as a cube there, is cubic only in a simple cubic lattice.

In order to prepare the way to more realistic calculations, we shall briefly review band theory and some key solid state physics technics which may not be familiar to the reader. We
shall then discuss some results for dripped neutrons in the inner crust, described within a mean field model.

\section{Inner layer of the crust}
\subsection{Mean field model}
In the absence of any previous calculation of neutron band structure except for a simple 1D model \cite{OyYam94}, even at the simplest level of approximation, we
shall adopt the single particle model of Oyamatsu and Yamada \cite{OyYam94}, to estimate the effective mass of the dripped neutrons in the bottom layers of the inner crust. Previously,
we only considered slab shaped and rod like nuclei for simplicity \cite{CCHI}. The effective mass was found to be very close to the bare mass. The reason was that the region we were focusing
on was nearly homogeneous. In the present work, we shall
extend the same analysis for lower densities where the crust is assumed to be composed of spherical nuclei arranged in a body centered cubic lattice. Simple cubic and face centered cubic structures
will also be considered for comparison. We shall also report results for the spherical and cylinder hole phases
near the crust-core transition layer.

\begin{figure}
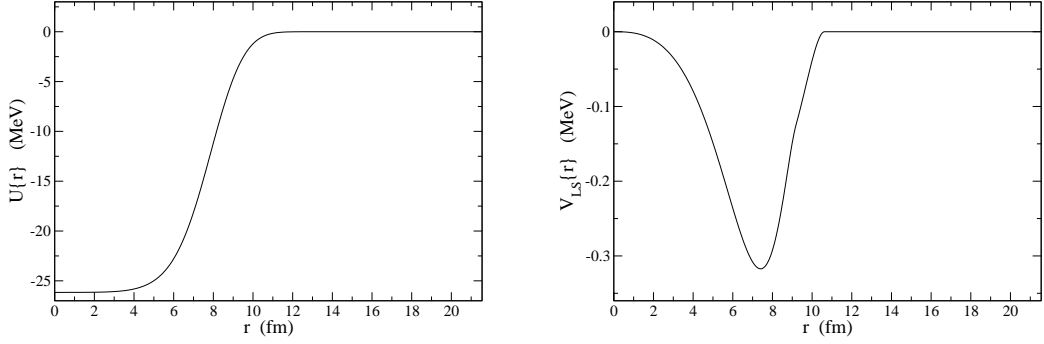

\centering
\epsfig{figure=oy_pot2.eps, height=4.5 cm}\hskip 1cm \epsfig{figure=oy_so_pot2.eps, height=4.5 cm}
\caption{Single particle central potential (left) and spin-orbit coupling potential (right) in the W-S sphere for the lowest baryon density considered {\bb \nn_{\rmb}=0.03\, {\rm fm}^{-3}\, \fb} from
model I of reference \cite{OyYam94}.}
\label{fig1bis}
\end{figure}

\begin{figure}
\centering
\epsfig{figure=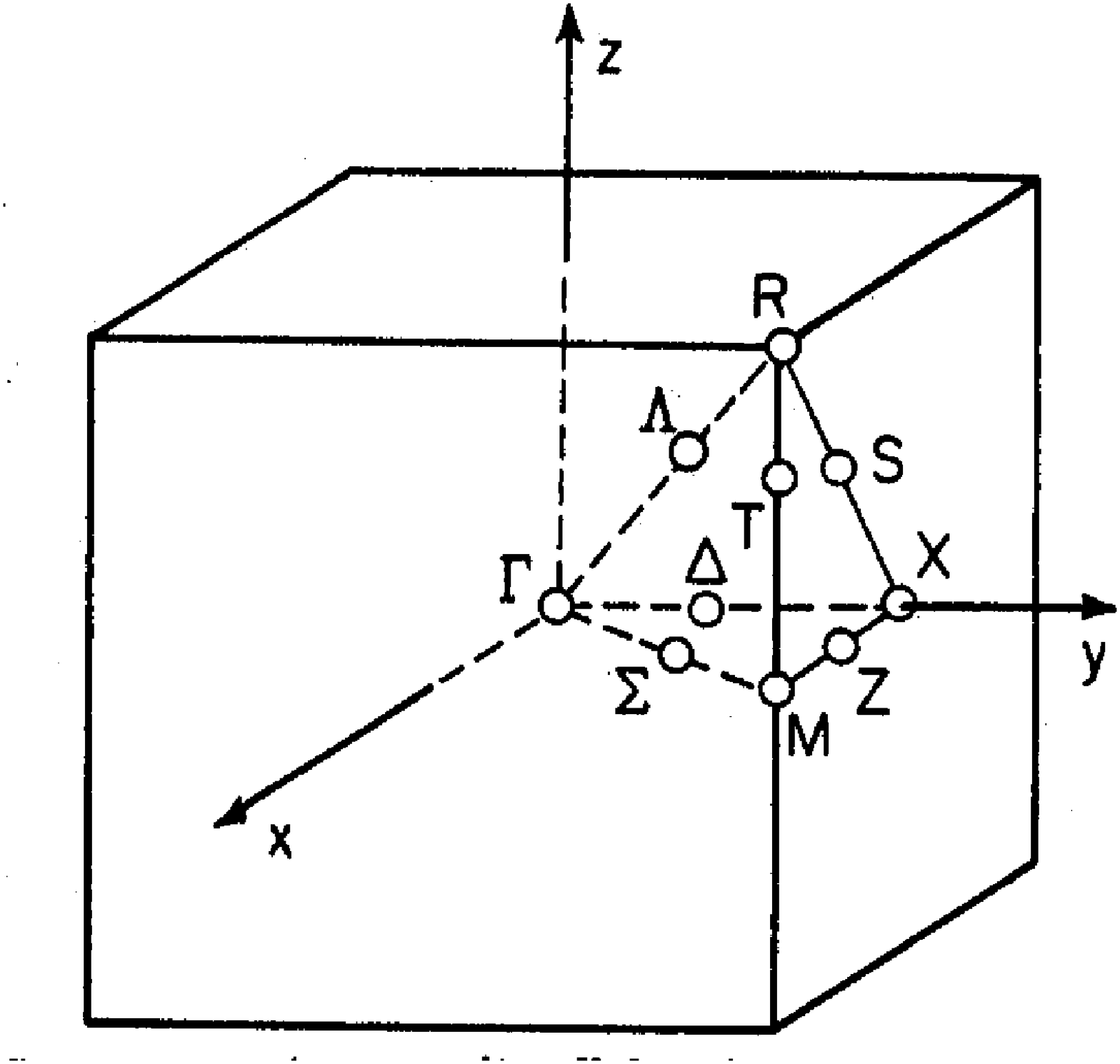, height=4.5 cm}
\epsfig{figure=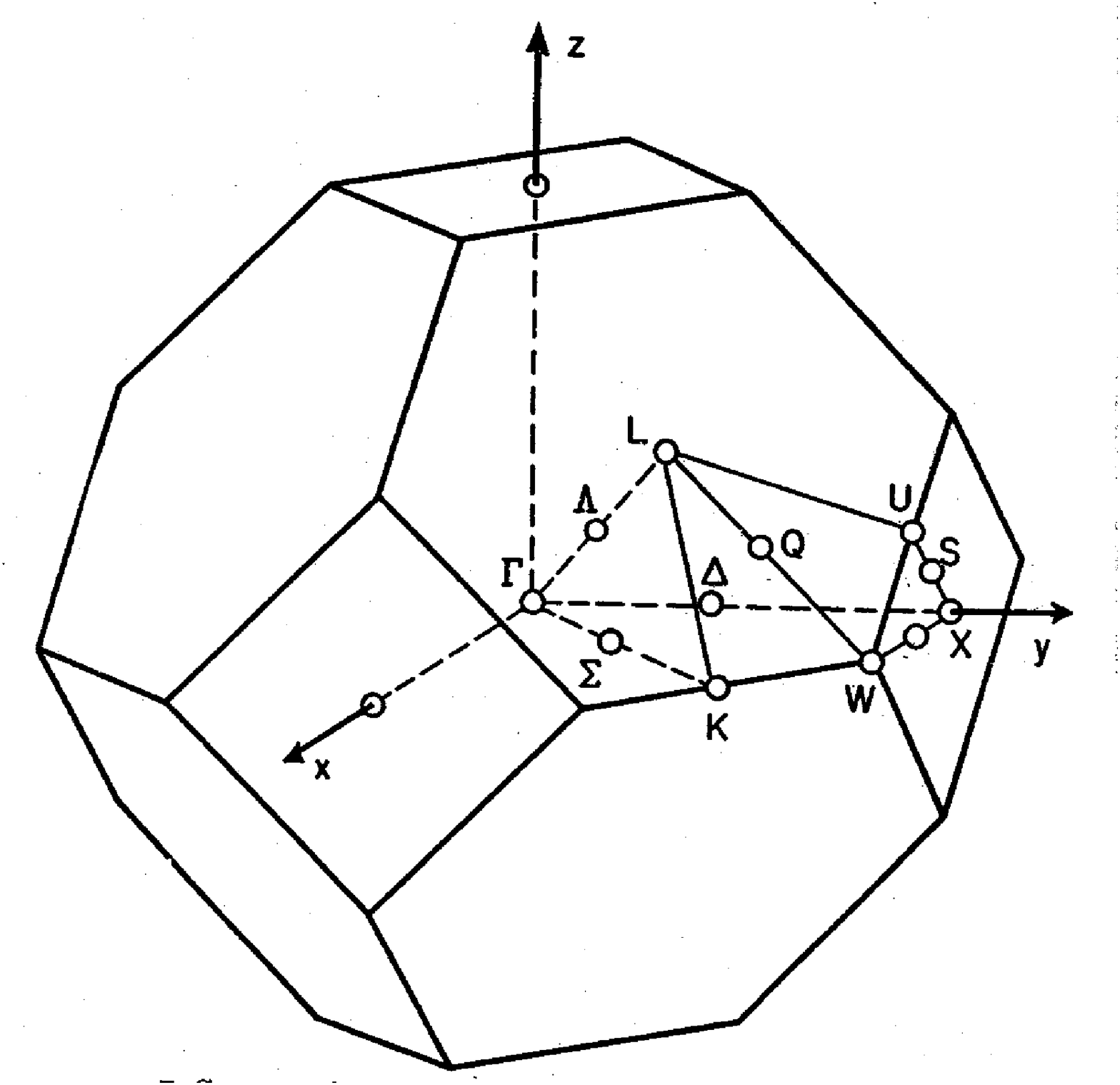, height=4.5 cm}
\epsfig{figure=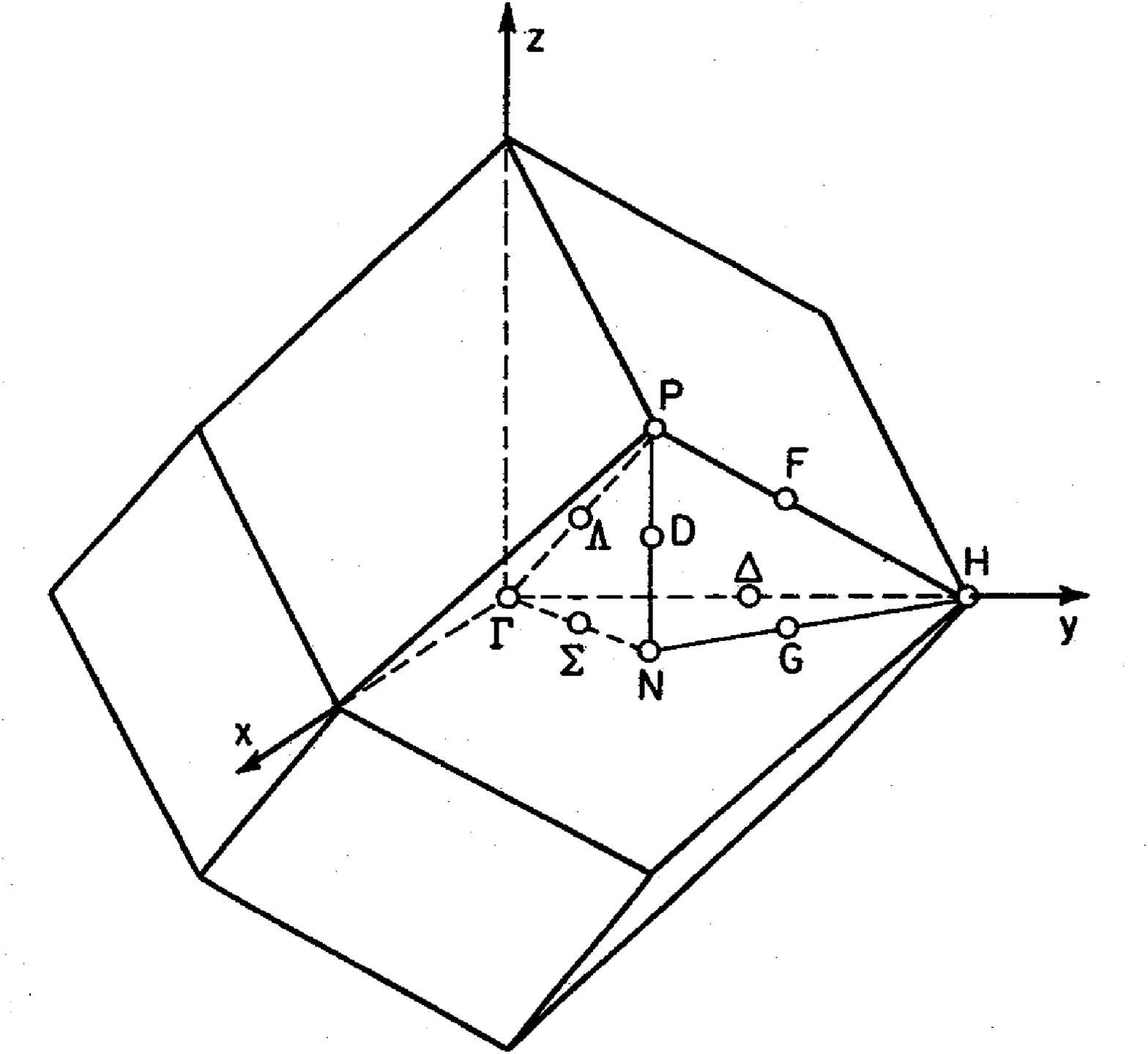, height=4.5 cm}
\caption{Wigner-Seitz cell of cubic lattices (from left to right) simple cubic, body centered cubic, face centered cubic with conventional labelling of high symmetry points and lines \cite{Koster57}.}
\label{fig1}
\end{figure}

In the model suggested by Oyamatsu et al \cite{OyYam94}, both bound and dripped neutrons are assumed to be
described as independent particles moving in a ``background'' phenomenological mean field. The neutron single particle states thus obey the
following Schr\"odinger equation
 {\be -\frac{\hba^2}{2 \mm}\Delta \phhi+\VV \phhi={\calE}\phhi\, . \label{Schrodeq}\fe} We further assume that nuclei are located
 at the nodes of an infinite crystal lattice. The single particle periodic potential {\bb \VV \fb} is thus constructed
from the potential {\bb U \fb} of Oyamatsu and Yamada given in the Wigner-Seitz sphere by the following procedure:
{\be \VV\{{\bf r}\}=\sum_{\bf \calR} U\{{\bf r}-{\bf \calR}\}\, , \label{periodic_pot}\fe}
where the sum goes over all lattice sites, i.e. {\bb {\bf \calR} = \sum_{\ita=1}^3\, \Bl^\ita {\bf \le}_\ita \fb}, where {\bb \Bl^\ita \fb} are integers and {\bb {\bf \le}_\ita \fb} basis lattice vectors.
The potential {\bb U \fb} was defined by Oyamatsu and Yamada as (equation (2.5) of reference \cite{OyYam94})
{\be U\{{\bf r}\} = \left(\frac{1}{\sqrt{\pi}\kappa}\right)^3\int{\rm d}^3 r^\prime U_0\{ {\bf r^\prime} \} e^{-|{\bf r} - {\bf r^\prime}|^2/\kappa^2} \, , \fe}
{\be U_0\{ {\bf r} \} = \frac{\delta v}{\delta \nn_n} \fe} where {\bb v \fb} is a potential energy density functional of homogeneous nuclear matter. The local neutron and proton densities
were obtained from a zero temperature Thomas-Fermi calculation in the W-S approximation. The energy density functional was parametrised such as to reproduce the properties of terrestrial nuclei
on the $\beta$ stability line and the equation of state of Friedman and Pandharipande for both pure neutron matter and symmetric nuclear matter. The parameter {\bb \kappa \fb} (as well as the spin-orbit
coupling parameters, see section \ref{res} below) was adjusted so as to give reasonable values for the single particle energies of {\bb ^{208}{\rm Pb} \fb}
(for further details, see \cite{Oyamatsu93, OyYam94}). We have used the parameter sets of model I from reference \cite{OyYam94}.
We have neglected the spin-orbit coupling introduced by Oyamatsu and Yamada since it is smaller by an order of magnitude compared to the central potential as shown on figure \ref{fig1bis}.
In order to take into account Bragg scattering, single particle states have to satisfy the Floquet-Bloch theorem  {\be \phhi_{\bf \kk }\{ {\bf r }+{\bf \calR } \}  = {\rm e}^{{\rm i}\,
{\bf \kk }\cdot {\bf \calR }}\phhi_{\bf \kk }\{ {\bf r } \} . \label{Blochthm}\fe}
As a consequence, the Schr\"odinger equation \rf{Schrodeq}\fr can be solved within any primitive cell \cite{Ashcroft81},
among which the Wigner-Seitz one possesses the full symmetry of the lattice, with boundary conditions given by \rf{Blochthm}\fr. The Wigner-Seitz (W-S)
cell is defined as the set of points that are closer to a given
lattice node than to any other. Exemples are shown for cubic crystals on figure \ref{fig1}. Except for simple cubic lattice, the W-S cell is a complicated polyhedron. In the work of Oyamatsu and Yamada, the lattice spacing was
defined as {\bb a_{\rm Oy}=\calV_{_{\rm cell}}^{1/3}, \fb}  where {\bb \calV_{_{\rm cell}} \fb}
is the volume of the W-S sphere. In the present work, we have defined the lattice spacing {\bb a \fb} for cubic lattices (cube length) such that the volume of the W-S polyhedron is equal to that of the sphere.

\subsection{Single particle energy spectrum}

 For each given {\bb \bf \kk \fb}, equation \rf{Schrodeq}\fr with boundary conditions \rf{Blochthm}\fr will admit a discrete set of eigenvalues {\bb \calE_\alp \fb} (indexed by
a Greek letter) referred as bands. Single particle energies will be ordered such that {\bb \calE_1<\calE_2<... \fb} which guarantees that for a given band, say {\bb \alp, \fb} the
energy {\bb \calE_\alp\{ {\bf \kk} \} \fb} will be continuous (besides it is analytic except for high symmetry points). Unlike the W-S approximation, the single particle energy spectrum in the Bloch case
 has a much more detailed structure.

It can be shown that the single particle energy is periodic in a ``reciprocal'' lattice \cite{Ashcroft81} whose nodes are given by
{\bb \bfK =\sum_{\ita=1}^3\,  \gl_\ita {\bf \lle}^{\ita}\fb}, where {\bb \gl_\ita \fb} are integers and the dual basis is defined by the dot products
{\be {\bf \lle^{\ita}\cdot \le_{\itb} }= 2\pi \delta^\ita_\itb\, .\label{dualbase} \fe} This means in particular that equation \rf{Schrodeq}\fr has only to be
solved for each momentum {\bb \bf \kk \fb}, within a domain known as the first Brillouin zone (BZ), i.e. the W-S cell of the reciprocal lattice. By considering rotational symmetry,
it can be shown that all the relevant information about the spectrum is contained within some ``irreducible'' domain of the BZ \cite{Kim99}. The solutions of Magierski et al
\cite{Magierski2001} are only those associated with the center of the first Brillouin zone,
namely {\bb \bf \kk=0 ,\fb} provided the lattice is simple cubic (for body centered or face centered cubic structures, their solutions are not of the Bloch type).

\begin{figure}
\centering
\epsfig{figure=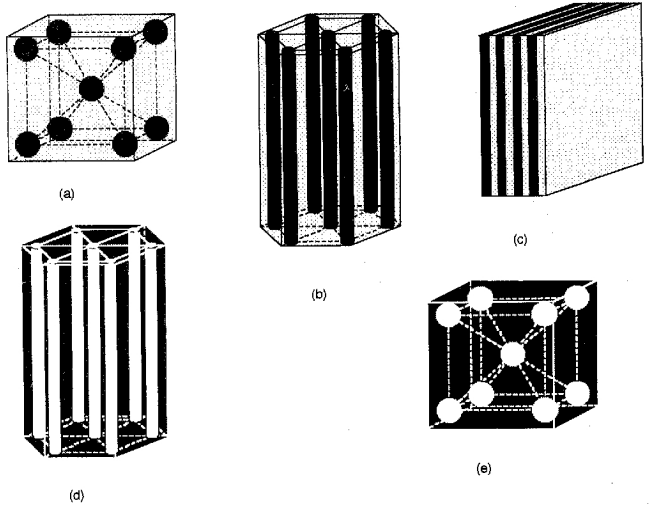, height=7 cm}
\caption{Nuclear configurations taken from reference \cite{Oyamatsu93}: (a) spherical nuclei in a body centered cubic lattice, (b) ``spaghettis'' in 2D hexagonal lattice, (c) ``lasagna'',
(d) ``antispaghetti'' in a 2D hexagonal lattice and (e) ``bubbles'' in a body centered cubic lattice.}
\label{fig2}
\end{figure}

We have solved the Schr\"odinger equation \rf{Schrodeq}\fr with Bloch type boundary conditions \rf{Blochthm}\fr variationally by a finite plane wave expansion
of the single particle wavefunctions, namely
{\be \phhi_{\bf \kk } \{ {\bf r } \} = \frac{1}{\sqrt{\calV_{\rm cell}}}
\sum_{\bfK } \widetilde{\phhi}_{\bf \kk }\{ {\bfK } \}\, {\rm e}^{ {\rm i}\,
({\bf \kk} + \bfK )\cdot{\bf r}}\, ,\label{Blochpacket}\fe} with an energy cutoff {\bb \calE_{\rm cut off}\fb} such that
{\be \frac{\hba^2 ({\bf \kk +\bfK })^2}{2 \mm} < \calE_{\rm cut off} \, .\fe}
The nuclear structures that we have considered are illustrated on figure \ref{fig2}.

\subsection{Effective mass}
The zero temperature many body ground state is merely obtained by filling all the single particle states whose energy {\bb {\calE}_\alp\{{\bf \kk}\} \fb} is
lower than the Fermi energy {\bb \mmu \fb}.
For each total neutron density {\bb \nn_{\rmn}, \fb} the Fermi energy {\bb \mmu\fb} is thus determined via the integral over the first Brillouin zone (BZ)
{\be \nn_{\rmn} =  \frac{g}{(2\pi)^3}\sum_\alp \int_{_{\rm BZ}}  {\rm d}^3 \kk\, \vartheta\{\mmu-{\calE}_\alp\{ {\bf \kk}\} \} \label{FermiEnergy}\fe} where
we have explicitely introduced the Fermi distribution by the Heaviside unit step distribution {\bb \vartheta\{x\}=1\fb} if {\bb x>1 \fb} and zero otherwise, allowing
each momentum state to be {\bb g \fb}-fold degenerate (here {\bb g=2 \fb} to account for the spin degeneracy). In particular,
 the ground state will be characterized by the Fermi surface bounding the occupied part of momentum space. In general the Fermi surface will
 consist of disconnected pieces, each piece being given by {\bb {\calE}_\alp\{ {\bf \kk} \}=\mmu .\fb}

 Introducing the dripped or free neutron density {\bb \nn \fb}, the effective mass (from which the entrainment coefficients can be obtained \cite{CCHII})
 is defined \cite{CCHI} by the simple formula
 {\be \mm_\star = \frac{\nn}{\calK},\label{EffMass}\fe}
 where the mobility scalar {\bb \calK \fb} is given by (summation over energy bands is implicit)
 {\be {\calK}=\frac{1}{3}\frac{g}{(2\pi)^3\hba}\oint_{_{\rm F}} \vv \,{{\rm d}\sS_{_{\rm F}}}, \label{Mobility}\fe }
 and the group velocity is defined by the usual expression
 {\be {\bf \vv} =\frac{1}{\hba} {\bf \nabla_{\kk}} \calE .\fe} This mobility scalar is related to the electric conductivity {\bb \sigm \fb} by the formula
 {\bb \sigm = e^2 \tau \calK \fb} in which {\bb e \fb} and {\bb \tau \fb} are the electric charge per particle and some characteristic relaxation time respectively.
 In terms of the effective mass, this formula reduces to the well known expression \cite{Ashcroft81}
 {\bb \sigma =  \nn e^2 \tau / \mm_\star \, . \fb}

 In the following sections, we have defined the dripped neutron states to be those states whose energy is positive {\bb {\calE}_\alp\{ {\bf \kk} \}>0 \fb} (the energy origin
 being taken as the largest possible value of the potential).

 \subsection{Wigner-Seitz approximation}

The W-S cell of body or face centered (unlike simple) cubic lattice is a nearly spherical polyhedron, this is the origin of the so called W-S
approximation \cite{WignerSeitz33}, according to which the W-S cell is taken as a sphere and more or less arbitrary boundary conditions are applied on the single particle wave functions. For instance, the prescription
given by Negele and Vautherin \cite{NV73} requires that on the sphere radius, radial wave functions with even orbital angular momentum quantum number {\bb l \fb} vanish and derivatives of the
radial wave functions with odd {\bb l \fb} vanish. Besides the density needs to be averaged in the vicinity of the cell edge in order to prevent the occurrence of unphysical density fluctutations. Within the current
framework, the Floquet-Bloch theorem (\ref{Blochthm}) ensures that the local total neutron density defined by
{\be \nn_{\rmn}\{ {\bf r} \} = \frac{g}{(2\pi)^3}\sum_\alp \int_{_{\rm BZ}}  {\rm d}^3 \kk\, |\phhi_{\alp\, \bf \kk} \{{\bf r} \}|^2 \vartheta\{\mmu-{\calE}_\alp\{ {\bf \kk}\} \} \label{density}\fe}
possesses the lattice symmetry, the single particle wavefunctions being normalized as
{\be \frac{1}{\calV_{_{\rm cell}}} \int_{_{\rm cell}}  {\rm d}^3 {\bf r}\, |\phhi_{\alp\, \bf \kk} \{{\bf r} \}|^2 =1.\fe}
This means in particular that the density gradient {\bb {\bf \nabla}\, \nn_{\rmn} \{ {\bf \calR} \} \fb} will vanish whenever the crystal is left invariant under space inversion, which means
that the density will be nearly constant (to first order) inside nuclei. In cases for which {\bb \bf r \fb} belongs to a symmetry plane, the density gradient will have no components orthogonal to this plane. This means
that the density profile will be essentially flat (to first order) around the cell boundary, along directions perpendicular to the W-S faces whenever the face is parallele to any symmetry plane. This is true for the simple cubic
and face centered cubic W-S cells. This is also true for the body centered cubic W-S cell excluding the hexagonal faces.

\section{Numerical methods}

\subsection{Brillouin zone integrations}

 Brillouin zone integrations are involved in many solid state physics calculations, as for instance in the particle density \rf{density}\fr. Analytic expressions for the integrand
 are usually unknown (except for a few academic cases), therefore one has to rely
 on numerical schemes based on discrete summations. Since the computation of energy bands is usually the most time consuming part of solid state calculations, several techniques have been developed in order to compute these integrals with the smallest number of terms. One of the most successful methods, which was
 pioneered by Baldereshi \cite{Baldereschi73} and subsequently improved by Chadi-Cohen \cite{ChadiCohen73}, is based on a weighted sum over a set of special symmetry points. The main idea relies on the fact that a
 function {\bb f\{{\bf \kk}\} \fb} which has the full symmetry of the reciprocal lattice,
 can be expanded over symmetrized plane waves (in the following we shall consider only cubic structures, for which there is no glide plane or screw symmetry axis \cite{Koster57})
 {\be f\{{\bf \kk}\} = \sum_s \tilde{f}_s\,  \frac{1}{|\cal P|} \sum_{\cal P} e^{{\rm i}\, {\cal P}\calR_s\cdot{\bf \kk}},\,\fe} where {\bb \cal P \fb} is any rotation of the lattice and {\bb |\cal P| \fb} their number
 (lattice vectors are ordered as {\bb |\calR_{s_1}| \geq |\calR_{s_2}|\fb} whenever {\bb s_1>s_2 \fb}).
 The integral of this function over the first Brillouin zone is rewritten as a sum over a set of unequivalent points {\bb {\bf \kk_j} \fb} with weight {\bb w\{{\bf \kk_j}\}: \fb}
 {\be \tilde{f}_{_0} = \frac{1}{\calV_{_{\rm BZ}}} \int_{_{\rm BZ}} f\{{\bf \kk}\} {\rm d}^3 \kk = \sum_{{\rm j}=1}^{\rm N} w\{{\bf \kk_j}\} f\{{\bf \kk_j}\} + R_{_{\rm N}}, \fe}
 {\be \sum_{{\rm j}=1}^{\rm N} w\{{\bf \kk_j}\} =1\fe}
 where the residual {\bb R_{_{\rm N}} \fb} is given by
 {\be R_{_{\rm N}} = - \sum_{s>0}  \tilde{f}_s \sum_{{\rm j}=1}^{\rm N} w\{{\bf \kk_j}\} \frac{1}{|\cal P|} \sum_{\cal P} e^{{\rm i}\, {\cal P}\calR_s\cdot{\bf \kk_j}}\, . \fe}
 By choosing the points {\bb \bf \kk_j \fb} such that all plane waves are exactely zero for all {\bb s \fb} up to some largest {\bb s_{\rm max} \fb}, the residual will contain
 only terms with lattice vectors {\bb |\calR|\geq |\calR_{s_{\rm max}}|\fb} for which the contribution will be vanishingly small provided the function {\bb  f\{{\bf \kk}\} \fb}  is sufficiently smooth.
 Eventually the integral is approximated by
 {\be  \frac{1}{\calV_{_{\rm BZ}}} \int_{_{\rm BZ}} f\{{\bf \kk}\} {\rm d}^3 \kk \simeq  \sum_{{\rm j}=1}^{\rm N} w\{{\bf \kk_j}\} f\{{\bf \kk_j}\} .\fe} The determination of these special points {\bb {\bf \kk_j} \fb}
 and their associated weights  {\bb w\{{\bf \kk_j}\} \fb} was rather complicated in the first scheme involving a recursion process. Several authors have provided formulae (see \cite{Watanabe92} and references
 therein).

 The convergence of the discrete sum with respect to the number of special points is exponential for smooth integrands and an
error of less than one per cent can be attained within a few points only. Unfortunately the presence of a sharp Fermi surface (especially in any metallic solid) introduces
discontinuities in the integrands, such as in equation \rf{density}\fr,
which spoil the convergence. This method is thereby most satisfactory for insulating materials and semiconductors. Nevertheless, accurate results can still be obtained with a small number of special points
by ``smearing'' the Fermi surface \cite{MethPax89}. Since the volume of the first Brillouin zone and of the Wigner-Seitz cell are related by
{\be \calV_{_{\rm BZ}} = \frac{(2\pi)^3}{\calV_{_{\rm cell}}} \label{duality},\fe} the number of special points for a given precision is smaller for larger cells. We have tested the special point method by
calculating the integral

\begin{figure}
\centering
\epsfig{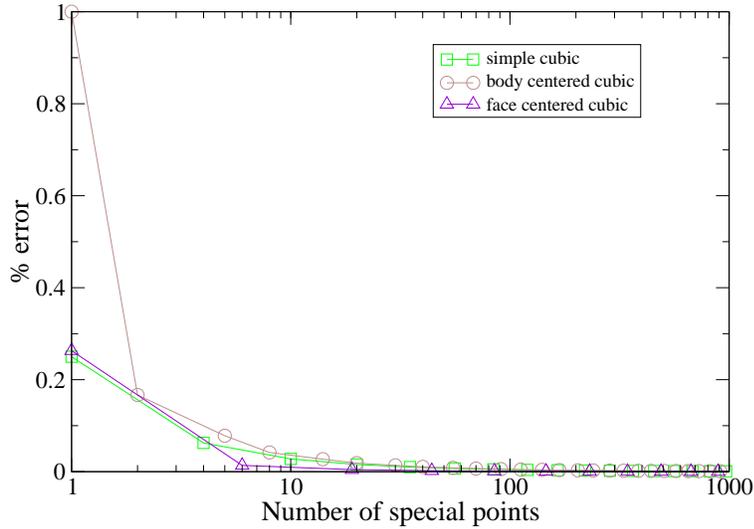}
\caption{Relative error in the computation of integrals (\ref{spe_test}) with respect to the number of points}
\label{fig3a}
\end{figure}

\begin{figure}
\centering
\epsfig{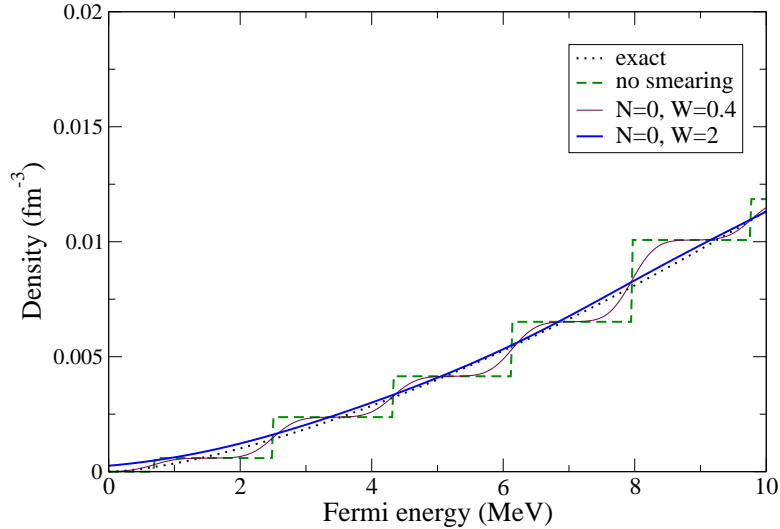}
\caption{Density as a function of the Fermi energy for the ``empty'' lattice model (body centered cubic lattice, lattice parameter=30 fm), with one single special point and with different smearing parameters
({\bb N \fb} is the number of approximants and {\bb W \fb} is the smearing width, see reference \cite{MethPax89} for the details)}
\label{fig3b}
\end{figure}

{\be \frac{1}{\calV_{_{\rm BZ}} } \int_{_{\rm BZ}}\, \kk^2 {\rm d}^3 \kk\, , \label{spe_test}\fe}
which is equal to {\bb 1/4, \, 3/8, \, 19/32 \fb} in units of {\bb (2\pi/a)^2 \fb} for simple cubic, body centered cubic and face centered cubic lattice respectively. The convergence is illustrated on figure \ref{fig3a}.
We have also shown on figure \ref{fig3b} how the smearing procedure of Methfessel et al \cite{MethPax89} could improve the convergence in the extreme case of one single special point, by computing the
total neutron density {\bb \nn_{\rmn} \fb}  as a function of the Fermi energy {\bb \mmu \fb} defined in equation \rf{FermiEnergy}\fr in the Shockley ``empty'' lattice test \cite{Shockley37}  for which we have
{\be \nn_{\rmn} = \frac{1}{3\pi^2}\left(\frac{\sqrt{2\mm \mmu}}{\hba}\right)^3 \fe}

\subsection{Fermi energy}
\label{fermienergy}

The determination of the Fermi energy is numerically expensive as it requires the calculation of all occupied single particle states (unlike the effective mass which depends only on the highest occupied states
on the Fermi surface). The computional cost is significant since the number of levels is of the order of several hundreds up to nearly one thousand. Nonetheless a useful estimate of the Fermi energy
with an accuracy of a few percents can still be easily obtained as follows. First, among all occupied states, one should distinguish
between core states whose wavefunction is localised in the neighborhood of nuclei and valence or conduction states whose wavefunction extends over all space. The core states are easily obtained by
finding the bound states of one isolated nucleus. Having found the number of such states, let's say {\bb N_{\rm core} \fb}, equation \rf{FermiEnergy}\fr can be written as

{\be \nn_{\rmn} = \frac{g}{(2\pi)^3}\calV_{_{\rm BZ}} N_{\rm core} + \frac{g}{(2\pi)^3}\sum_\alpha\int_{_{\rm BZ}}  {\rm d}^3 k\, \vartheta\{ \mmu-{\calE}_\alp\{ {\bf k}\} \} \vartheta\{ {\calE}_\alpha\{ {\bf k} \}-\calE_{\rm core}\}, \fe}
where {\bb \calE_{\rm core} \fb} is the energy of the highest core state. The determination of the Fermi energy is essentially a state counting procedure which becomes less and less sensitive to the detailed structure
of the energy spectrum as the number of particles is increased. Consequently assuming that the valence states can be roughly described as non interacting particles, the previous equation thus leads to the following estimate:

{\be \mmu \simeq \calE_{\rm core} + \frac{\hba^2}{2\mm}\biggl( (\nn_{\rmn}-\nn_{\rm core})\frac{6 \pi^2}{g} \biggr)^{2/3} ,\fe} where from \rf{duality}\fr we have {\bb \nn_{\rm core} = g N_{\rm core}/\calV_{\rm cell} .\fb}

For instance, for the baryon density {\bb \nn_{\rmb}=0.03\, {\rm fm}^{-3} ,\fb} we found for the core states: three s states, two p states, two d states, one f state and one g state. The total number of
core states is thus equal to {\bb N_{\rm core}=35 \fb} and the Fermi energy is approximately given by {\bb \mmu \simeq 18.02 \fb} MeV, which is about {\bb 5\% \fb} larger than the value
{\bb \mmu = 17.16 \fb} MeV, obtained by direct integrations of \rf{FermiEnergy}\fr via special points. The value of the Fermi energy was also found to be completely unsensitive to the lattice structure,
which is due to the fact that equation \rf{FermiEnergy}\fr does not explicitely depend on the precise values of energy bands but only on their number in a given energy range.

\subsection{Fermi surface integrations}

Calculations of the effective mass \rf{EffMass}\fr require a fine mesh sampling of the Fermi surface. One of the most accurate methods for performing Fermi surface integrations is based on the Gilat-Raubenheimer (GR)
scheme \cite{Janak71}. The integration over the Fermi surface is first reduced by symmetry
to the irreducible wedge of the first Brillouin zone which is partitionned into microcells. The integral is thereby discretized as a sum over the microcells with suitable weight factors to account for the fact that
microcells may overfill each wedge. The Fermi surface is then approximated by a plane inside the microcells which it intersects, so that the integration within each cell can be calculated analytically \cite{Gilat81}.

We have implemented the
original linear extrapolation method with cubic cells \cite{Gilat66}. The energy is
linearly extrapolated from the values of the energy and
velocity at the center of the cell. This scheme therefore requires the energy bands as well as their gradients, whose evaluation with plane waves is straightforward via the Hellmann-Feynman theorem \cite{Feynman39},
namely
{\be {\bf \vv} = \frac{\hba {\bf \kk }}{\mm} +\sum_\bfK \frac{\hba \bfK}
{\mm} |\widetilde{\phhi}_{\bf \kk }({\bfK})|^2\, ,\label{groupvel}\fe} with the normalisation of the wavefunction given by
{\be \sum_{\bfK } |\widetilde{\phhi}_{\bf \kk }({\bfK })|^2 =1\, .\fe}
The GR method seems to be not so popular as the tetrahedron method in solid state physics, in which the irreducible domain is divided into tetrahedra and the energy is linearly interpolated from its values at the four
corners of each tetrahedron \cite{Lehmann7072}\cite{Jepsen71}. One of the main reasons is that the calculation of the velocity may be computationally
 expensive with more elaborate basis functions than plane waves.
 Nevertheless the linear extrapolation of the energy within the GR method, which is a first order Taylor expansion around the cell center, is
 a better approximation than an interpolation. Since in the present case the integrand in equation \rf{Mobility}\fr involves the energy gradient, the GR method seems much more appropriate. Besides interpolation
 faces the band crossing problem: the ordering of the energy bands
 with increasing energy ensures that the energy is continuous for a given band but does not prevent discontinuities in the energy gradient from occuring. Unlike extrapolation, interpolation will yield an incorrectly
 small gradient resulting in a loss of convergence (see \cite{Gilat72}\cite{Gilat75}\cite{Gilat72b} for a discussion about systematic errors).

\begin{figure}
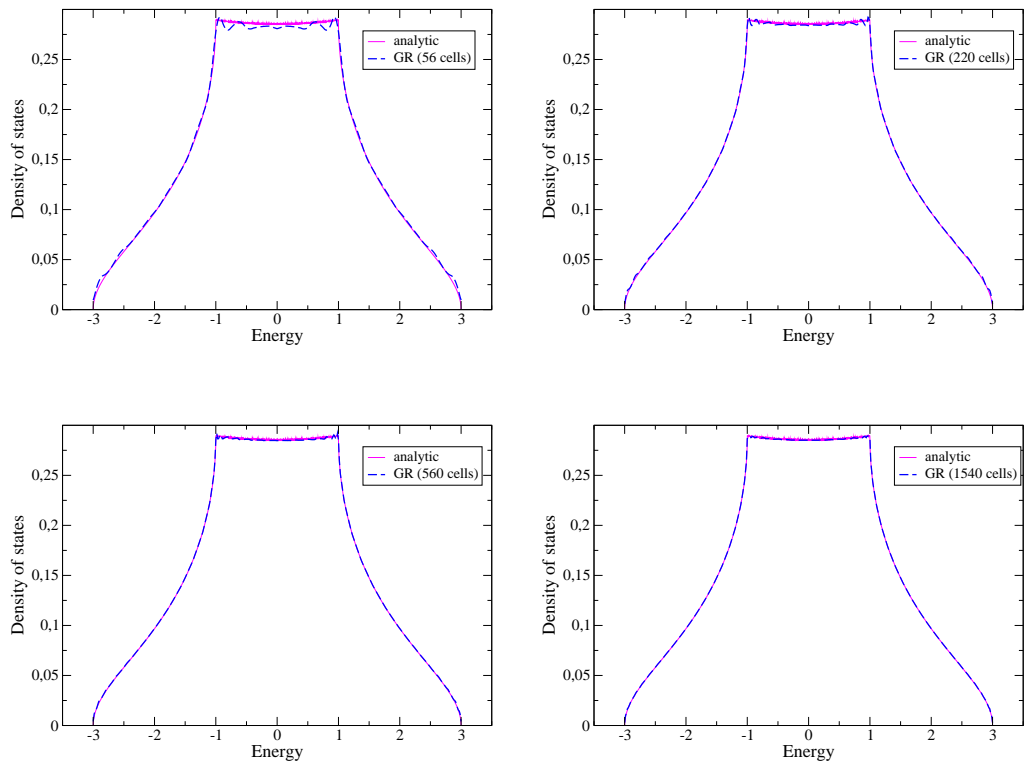

\centering
\epsfig{figure=cubium_dos1.eps, height=4.5 cm} \hskip 0.5 cm
 \epsfig{figure=cubium_dos2.eps, height=4.5 cm} \vskip 1 cm
 \epsfig{figure=cubium_dos3.eps, height=4.5 cm} \hskip 0.5 cm
 \epsfig{figure=cubium_dos4.eps, height=4.5 cm}
\caption{Density of states of the cubium model computed with the GR method and compared with the analytic expressions given by Morita et al \cite{Morita71}}
\label{fig3}
\end{figure}

 In order to test this scheme, we have employed the  ``cubium'' toy model, {\it i.e.} a simple cubic crystal with one single {\it s} band
 whose momentum dispersion in cartesian coordinates is given by {\bb {\calE}\{ {\bf \kk} \}= - (\cos \kk_x  +\cos \kk_y + \cos \kk_z ). \fb} Analytic expressions for the density of single particle states
 at energy {\bb\calE\fb}, defined by
 {\be \calN\{\calE\} = \frac{{\rm d}\nn}{{\rm d}\calE}=\frac{g}{(2\pi)^3\hba} \oint \,\frac{{\rm d}\sS_{_\calE}}{\vv}. \fe\label{Dos}}
 are available in terms of the elliptic integral of the first kind \cite{Morita71}. As can be seen on figure \ref{fig3} ({\bb g=2 \fb}) the GR method yields results in very good agreement with the analytic
 expressions, in particular in the vicinity of the critical points associated with {\bb \vv=0 \fb} (producing the ``kinks'' in the density of states).

\section{Results}
\label{res}

\begin{figure}
\centering
\epsfig{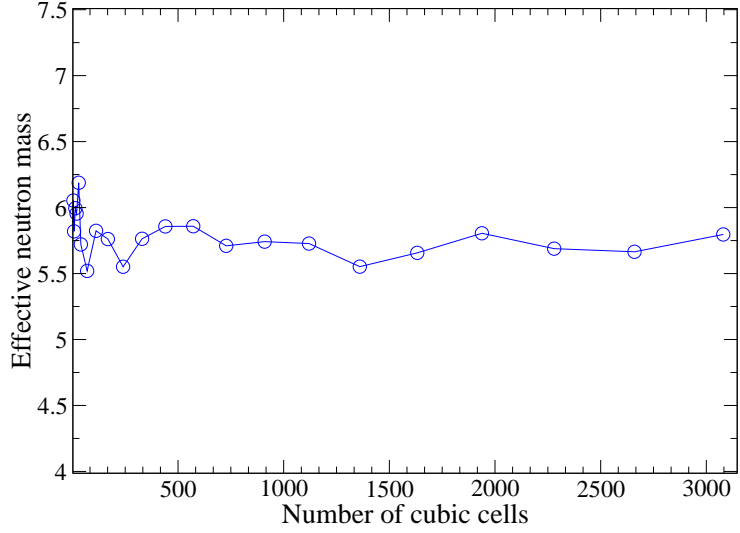}
\caption{Convergence of the effective neutron mass {\bb \mm_\star/\mm \fb} with respect to the number of GR cells, for baryon density {\bb \nn_{\rmb}=0.055\, {\rm fm}^{-3} \fb}
with nuclei in a body centered cubic lattice}
\label{fig5a}
\end{figure}

\begin{figure}
\centering
\epsfig{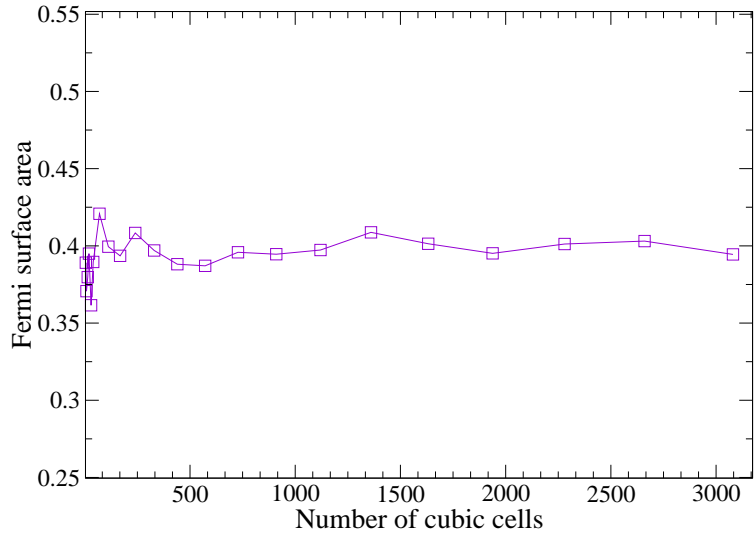}
\caption{Convergence of the Fermi surface area {\bb \sS/\sS_{_{\rm gas}} \fb} with respect to the number of GR cells, for baryon density {\bb \nn_{\rmb}=0.055\, {\rm fm}^{-3} \fb}
with nuclei in a body centered cubic lattice}
\label{fig5b}
\end{figure}

We have considered the three types of cubic lattices, namely simple cubic, face centered cubic and body centered cubic for spherical nuclei and bubbles ({\it i.e.} periodic lattice of ``holes'' containing dripped neutrons
 in uniform nuclear matter). The Fermi surface integrations have been carried out with of the order of {\bb \sim 10^3 \fb} cubic cells, for which the relative deviation was a few percents as shown on figures
 \ref{fig5a} and \ref{fig5b}.

The effective mass {\bb \mm_\star \fb} shown on
figure \ref{fig6b} has been found to be greatly enhanced due to Bragg scattering, reaching {\bb \mm_\star \sim 15\, \mm \fb} at a total baryon density {\bb \nn_{\rmb}=0.03\, {\rm fm}^{-3} .\fb} Deeper
inside the crust, the renormalization is smaller, tending to {\bb \mm_\star \sim \mm \fb} at very high density as the single particle potential tends to a constant. We have also shown on figure \ref{fig7} the
transverse effective neutron mass in the ``exotic'' phases \cite{CCHI} including the ``antispaghetti'' layer, where non spherical nuclei occur (only the velocity components respectively perpendicular to the slabs or
to the rods, for the ``lasagna'' or ``(anti)spaghetti'' phases are included in equation \rf{Mobility}\fr for the mobility scalar ; the other velocity components have the same expressions as those of a noninteracting gas
and therefore the corresponding effective mass coincide with the ordinary mass). The dependence on the lattice structure is rather weak. Numerical results are summarized in appendix \ref{results}.
We have also carried out the calculations with the estimate of the Fermi energy suggested in section \ref{fermienergy}. While this is usually a good approximation for bulk quantities such as the conduction neutron
density, the errors are much larger for quantities depending only on states at the Fermi level. In particular, the resulting effective mass is about {\bb 20 \% \fb} lower at baryon density
{\bb \nn_{\rmb} =0.03\, {\rm fm}^{-3} \, , \fb} having value {\bb \mm_\star/\mm \simeq 14 \, ,\fb} where the relative error in the conduction neutron density is less than on percent.

\begin{figure}
\centering
\epsfig{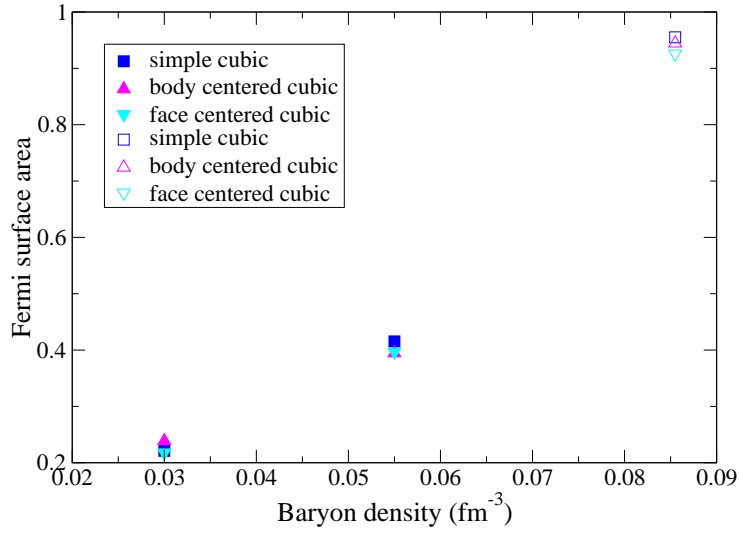}
\caption{Neutron Fermi surface area $\sS_{_{\rm F}}/\sS_{_{\rm gas}}$ for configurations with spherical nuclei (filled symbols)  and bubbles (unfilled symbols) in cubic lattices}
\label{fig6a}
\end{figure}

\begin{figure}
\centering
\epsfig{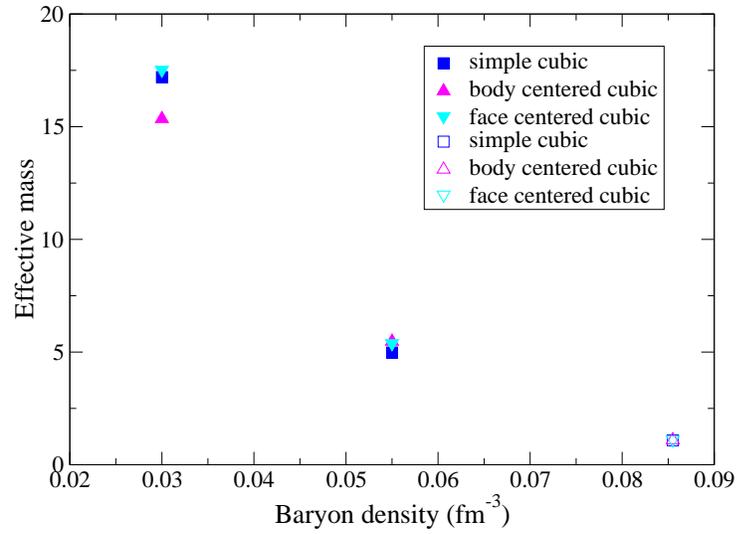}
\caption{Effective neutron mass $\mm_\star/\mm$ for configurations with spherical nuclei (filled symbols)  and bubbles (unfilled symbols) in cubic lattices}
\label{fig6b}
\end{figure}

\begin{figure}
\centering
\epsfig{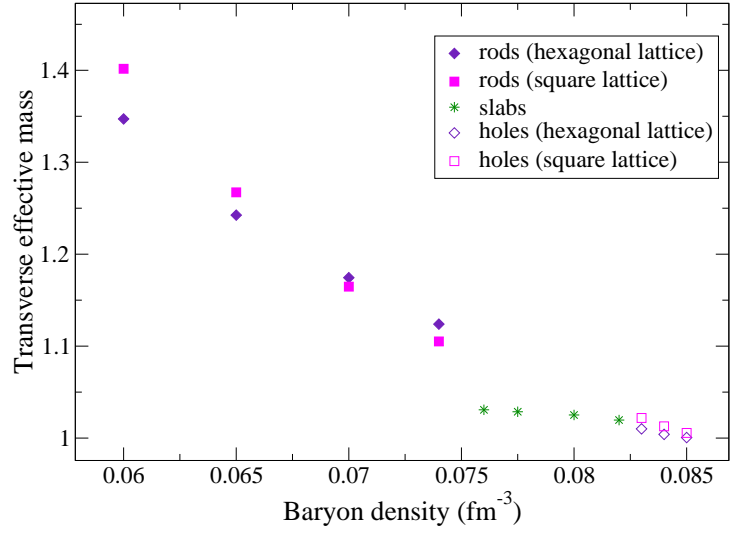}
\caption{Transverse effective neutron mass $\mm^\perp_\star/\mm$ for ``exotic'' configurations}
\label{fig7}
\end{figure}

In order to understand the origin of this effective mass enhancement, we have compared the Fermi surface area {\bb \sS_{_{\rm F}} \fb} with its expression obtained from
the assumption that the dripped neutrons form an ideal Fermi gas of density {\bb \nn\, , \fb} which is given by
{\be \sS_{_{\rm gas}} = 4\pi g (6\pi^2 \nn/g)^{2/3} \label{Sgas} \, . \fe}
We have found that the Fermi surface area is strongly reduced in the outer layers compared to that of the sphere, whereas it tends to its uniform
expression \rf{Sgas}\fr near the crust-core interface as illustrated on figure \ref{fig6a}. It is first to be remarked that the ``enclosed''
Fermi volume depends only on the density, as is easily seen on equation \rf{FermiEnergy}\fr and therefore does not depend on the particular shape of the Fermi surface.
It is well known that among all the \emph{closed} surfaces of a given volume, the sphere has the minimum surface area. In the present case however, the opening of numerous
band gaps in the single particle energy spectrum leads to a Fermi surface composed of \emph{disjoint open} pieces. By band gaps, we mean avoided crossings of enery band sheets in
{\bb \kk \fb} space. This is the reason why the Fermi surface area is found to be smaller than that of the
sphere in the low density layers of the crust where the gaps take their largest values.

It is quite instructive to evaluate the mean Fermi velocity {\bb {\bar \vv}_{_{\rm F}} \fb} defined by
{\be {\bar \vv}_{_{\rm F}} = \frac{1}{\sS_{_{\rm F}}} \oint_{_{\rm F}} \vv \,{{\rm d}\sS_{_{\rm F}}}.\fe} It is readily seen that it is expressible, in terms of the Fermi surface area {\bb \sS_{_{\rm gas}} \fb}
and Fermi velocity {\bb \vv_{_{\rm gas}} \fb} of the non interacting neutron gas, as
{\be \frac{{\bar \vv}_{_{\rm F}}}{\vv_{_{\rm gas}}} = \frac{\sS_{_{\rm gas}}}{\sS_{_{\rm F}}}\frac{\mm}{\mm_\star}\, .\fe} For instance, at the lowest density
we considered, namely {\bb \nn_{\rmb}=0.03\, {\rm fm}^{-3}, \fb} the mean Fermi velocity is about one third the Fermi velocity of the gas. The ratio increases to one half at {\bb \nn_{\rmb} =0.055\, {\rm fm}^{-3} \fb}
and tends to one as the density approaches the crust-core transition density. This is a clear and intuitive manifestation of Bragg scattering.

The spin-orbit coupling, as we have seen, is very small compared to the central potential in the bottom layers we have considered (see figure \ref{fig1bis}). However such a term would probably
not be negligible in the lower density regions of neutron star crust. The presence of this term in the Schr\"odinger equation \rf{Schrodeq}\fr preserves the lattice symmetry which implies that single
particles wave functions can still be expressed as Bloch states. However the spin-orbit coupling breaks the spin symmetry and therefore raise some degeneracy. As a result, this will
open new gaps in the energy spectrum, in the vicinity of which the group velocity will be vanishingly small. Consequently, it can be seen from \rf{Mobility}\fr and \rf{EffMass}\fr that
the resulting effective masses would be expected to be increased by the spin-orbit coupling since the density would not be much affected. This conclusion is confirmed by a numerical calculation in
the ``spaghetti'' phase. The computations are quite involved since the size of the secular (complex) matrices is doubled. We have taken into  account the spin-orbit coupling given by Oyamatsu and
Yamada (equation (2.12) of reference \cite{OyYam94}), which is given in the W-S approximate
cell (a cylinder in this case directed along the  {\bb z \fb} axis) by an expression of the form
  {\be \VV_{\rm LS} \{ r\} = \frac{1}{r}\left(\lambda_1 \frac{{\rm d}\nn_{\rmb}}{{\rm d} r}-\lambda_2 \frac{\rm d}{{\rm d}r}(\nn_{\rmn} - \nn_{\rmp})\right)\frac{1}{2} l_z \sigma_z. \fe}
 where {\bb r \fb} is the distance to the cylinder axis, {\bb l_z \fb} the angular momentum component along the cylinder axis
 and {\bb \sigma_z\fb} the diagonal Pauli spin matrix. The parameters {\bb \lambda_1 \fb} and {\bb \lambda_2 \fb} were adjusted so as to give the correct sequence of single particles states of
 {\bb ^{208}{\rm Pb}\, . \fb} We have used the values of model I from reference \cite{OyYam94} namely {\bb \lambda_1 =175.8\, \rm MeV.fm^5 \fb} and  {\bb \lambda_2 =16.39\, \rm MeV.fm^5 \, . \fb}
 We have carried out the calculation at the total baryon density {\bb \nn_{\rmb} = 0.06\, {\rm fm}^{-3} \fb} for the hexagonal lattice. Since the densities varies
 smoothly, the spin-orbit coupling has a small effect, yet the opening of band gaps can be clearly observed as shown on figure \ref{fig9}. The transverse
 effective mass is found to be slightly larger with than without spin-orbit coupling as expected, respectively {\bb \mm^\perp_{\star}/\mm=1.37 \fb} and {\bb \mm^\perp_{\star}/\mm=1.35 \fb} while
 the Fermi surface area is smaller, respectively  {\bb \sS_{_{\rm F}}/\sS_{_{\rm gas}} = 0.88 \fb} compared to {\bb \sS_{_{\rm F}}/\sS_{_{\rm gas}} = 0.89\, . \fb}

\begin{figure}
\centering
\epsfig{figure = 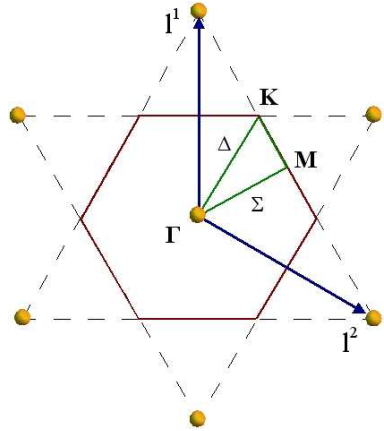, width=5 cm}
\caption{First Brillouin zone and irreducible domain of an hexagonal lattice with the conventional
labeling of high symmetry points and lines.
\label{fig8}}
\end{figure}

\begin{figure}
\centering
\epsfig{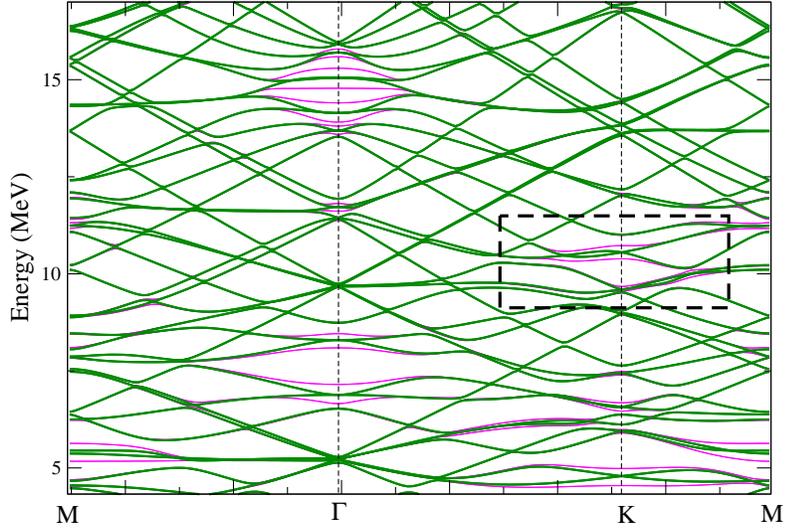}
\caption{Neutron band structure, with (thin lines) and without (thick lines) along high symmetry directions shown on figure \ref{fig8}
for an hexagonal lattice of spaghetti (rod) like nuclei at baryon density {\bb \nn_{\rmb} =0.055\, {\rm fm}^{-3} \, .\fb}} 
\label{fig9}
\end{figure}

\begin{figure}
\centering
\epsfig{figure=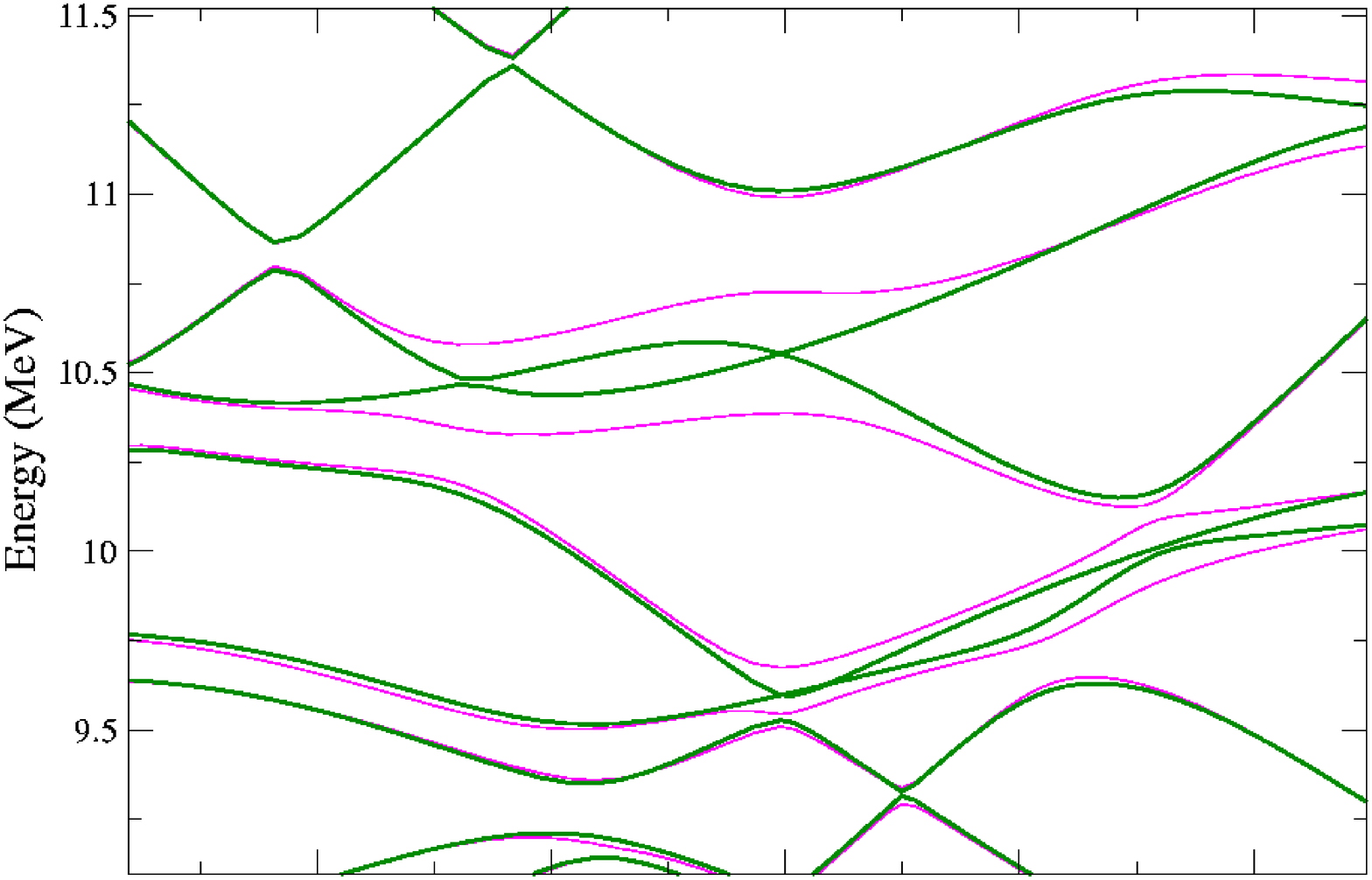, height=8 cm}
\caption{Zoom of the region in the box of figure \ref{fig9}. Band splittings induced by spin-orbit coupling can be clearly seen.}
\label{fig10}
\end{figure}

\bigskip
\section{Conclusion}
We have considered the band effects in neutron star crust induced by Bragg scattering of dripped neutrons upon the solid lattice of nuclei. This is the analog in a nuclear context,
of a well known phenomenon in ordinary solid state physics whereby valence electrons are being scattered by ions leading notably to an understanding of semiconductor band gaps. Based on
a mean field model with Bloch type boundary conditions, we have computed the effective dripped neutron macroscopic mass {\bb \mm_\star \fb} necessary for evaluating the entrainment coefficients in hydrodynamical
descriptions of neutron stars \cite{CCHI, CCHII}. We found that this effective mass is increased by several hundred percents compared to the ordinary neutron mass in the middle layers of the inner crust but becomes
negligible near the crust-core boundary where nuclei merge into a uniform mixture of neutrons, electrons and protons (and perhaps of some other
mesons). We have also found that the lattice structure plays a minor role. However the differences are
significant between 3D crystals and the 1D or 2D liquid crystal configurations that were previously studied within the same model (respectively slab shaped and rod like nuclei).

The present results also suggest that the widely used Wigner-Seitz approximation, whereby Bragg scattering is neglected, may be questionable for the study of neutron star crust.
In particular, it is to be remarked that the very large values of the effective mass {\bb \mm_\star \fb} that we found in the middle layers tend to indicate that dripped neutron shell (actually band)
effects may have significant consequences for the neutron star crust equilibrium structure compared to that obtained in the W-S approximation. It has been recently shown that the composition
of the inner crust of neutron star is quite sensitive to the proton shell effects \cite{Dutta2004}. The unbound neutron shell effects have been investigated
by Magierski and Heenen \cite{Magierski2001} who suggested that such quantum effects play an important role in the determination of the neutron star crust
equilibrium structure since the energy difference
between different nuclear shapes, lattice structures and compositions is typically small. Their analysis was based on a semiclassical approach. They have recently
carried out Skyrme Hartree-Fock calculations \cite{Magier2003} but with ordinary periodic boundary conditions. The implementation
of Bloch type boundary conditions within more realistic models, especially taking into account neutron pairing as outlined recently \cite{CCH2}, should therefore be further investigated.

\bigskip
{\bf Acknowledgements}
\medskip

The author would like to thank B. Carter, P. Haensel and D. G. Yakovlev for instructive conversations.

\vfill\eject
{\bf APPENDIX}

\appendix
\section{Numerical results}
\label{results}

Effective dripped neutron mass {\bb \mm_\star \fb} for spherical nuclei and bubbles in a body centered cubic lattice:

\bigskip

\begin{tabular}{|p{2cm} || p{2cm} | p{2cm} | p{2cm} |  }\hline
{\bb \nn_{\rmb}\, ({\rm fm}^{-3}) \fb} & {\bb \nn_{\rmn}\, ({\rm fm}^{-3}) \fb}  & {\bb \nn/\nn_{\rmn} \fb} & {\bb \mm_{\star}/\mm \fb}  \\ \hline
0.03 &  0.029 & 0.94 & 15.4  \\ \hline
0.055 & 0.053  & 0.95 & 5.5  \\ \hline \hline
0.0855 & 0.0826 & 0.98 & 1.07 \\ \hline
\end{tabular}

\bigskip

\noindent Transverse effective dripped neutron mass for ``pasta'' phases: ``spaghettis'' (hexagonal lattice), ``lasagna'' and cylinder holes (hexagonal lattice):

\bigskip

\begin{tabular}{|p{2cm} || p{2cm} | p{2cm} | p{2cm} |  }\hline
{\bb \nn_{\rmb}\, ({\rm fm}^{-3}) \fb} & {\bb \nn_{\rmn}\, ({\rm fm}^{-3}) \fb}  & {\bb \nn/\nn_{\rmn} \fb} & {\bb \mm^\perp_{\star}/\mm \fb}  \\ \hline
0.06 & 0.0581 & 0.9553 & 1.347   \\ \hline
0.065 & 0.063  & 0.9578 & 1.242\\ \hline
0.07 & 0.0678  & 0.9606 & 1.174  \\ \hline
0.074 & 0.0716  & 0.9633 & 1.124 \\ \hline\hline
0.076  & 0.0735 & 0.9634 & 1.031  \\ \hline
0.0775 & 0.0749 &  0.9646 & 1.029  \\ \hline
0.08  & 0.0773 &  0.9666 & 1.025   \\ \hline
0.082 & 0.0792  & 0.9687 & 1.020  \\ \hline\hline
0.083 & 0.0802  & 0.9734 & 1.010   \\ \hline
0.084 & 0.0811  & 0.9746 & 1.003  \\ \hline
0.085  & 0.0821 & 0.9762 & 1.001  \\ \hline
\end{tabular}


\bigskip

\begin{thebibliography}{99}

\bibitem{CCHI} B. Carter, N. Chamel \& P. Haensel,
``Entrainment coefficient and effective mass for conduction neutrons in neutron star crust: simple microscopic models'',
to appear in Nucl. Phys. A [nucl-th/0402057]

 \bibitem{CCHII} B. Carter, N. Chamel \& P. Haensel,
``Entrainment coefficient and effective mass for conduction neutrons in neutron star crust: Macroscopic treatment'',
preprint [astro-ph/0408083]

\bibitem{AB76} A. F. Andreev, E. P. Bashkin,
``Three velocity hydrodynamics of superfluid solutions'',
{\it Sov. Phys. JETP} {\bf 42} (1976) 164-167.

\bibitem{PR95} C. J. Pethick, D. G. Ravenhall,
``Matter at large neutron excess and the physics of neutron star crusts'',
{\it Annu. Rev. Part. Sci.} {\bf 45} (1995) 429-484.

\bibitem{Yak2001} D. G. Yakovlev, A. D. Kaminker, O. Y. Gnedin, P. Haensel,
``Neutrino emission from neutron stars'',
{\it Phys.  Rep. } {\bf V354, I1-2} (2001), 1-155.

\bibitem{Magierski2001} P. Magierski \& P. H. Heenen,
``Structure of the inner crust of neutron stars: crystal lattice or disorded phase?'',
{\it Phys. Rev. C }{\bf V65} (2002) 045804, 1-13.

\bibitem{MMM2004} F. Montani, C. May, H. M\"uther,
``Mean field and pairing properties in the crust of neutron stars'',
[nucl-th/0401024]

\bibitem{OyYam94} K. Oyamatsu, M. Yamada,
``Shell energies of non-spherical nuclei in the inner crust of
a neutron star'',
{\it Nucl. Phys.} {\bf A578} (1994), 181-203.

\bibitem{CCH2} B. Carter, N. Chamel \& P. Haensel,
``Effect of BCS pairing on entrainment in superfluid current of neutrons in neutron star crust'',
preprint [astro-ph/0406228]

 \bibitem{BJK96} M. Borumand, R. Joynt, W. Kluzniak,
``superfluid densities in neutron star matter'',
{\it Phys. Rev. C} {\bf V54-N5} (1996) 2745-2750.

\bibitem{CJ2003} G. L. Comer, R. Joynt,
``A relativistic mean field model for entrainment in general relativistic superfluid neutron stars'',
{\it Phys. Rev. D}{ \bf  68} (2003), 023002
[gr-qc/0212083]

\bibitem{NV73} J. W. Negele, D. Vautherin,
``Neutron star matter at sub nuclear densities'',
{\it Nucl. Phys.} {\bf A207} (1973), 298-320.

 \bibitem{Ashcroft81} Ashcroft \& Mermin,
{\it Solid State Physics}
(Holt-Saunders International Editions, 1981)

  \bibitem{Kim99} S. K. Kim,
 {\it Group theoretical methods and applications to molecules and crystals}
 (Cambridge University Press, 1999)

\bibitem{WignerSeitz33} E. P. Wigner \& F. Seitz ``On the constitution of
metallic sodium'',
{\it Phys. Rev. } {\bf V43-I10}, (1933) 804-810 ;{\it Phys. Rev. }
{\bf V46-I6}, (1934) 509-524.

\bibitem{Baldereschi73} A. Baldereschi,
 ``Mean value point in the Brillouin zone'',
 {\it Phys. Rev. B} {\bf V7-N12} (1973), 5212-5215.

\bibitem{ChadiCohen73} D. J. Chadi, M. L. Cohen,
 ``Special points in the Brillouin zone'',
 {\it Phys. Rev. B} {\bf V8-N12} (1973), 5747-5753.

\bibitem{Watanabe92} J. Hama \& M. Watanabe, ``General formulae for the special points and their weighting factors in k-space integration''
 {\it J. Phys.: Condens. Matter}{\bf 4} (1992), 4583-4594.

\bibitem{MethPax89} M. Methfessel \& A. T. Paxton, ``High-precision sampling for Brillouin zone integration in metals''
 {\it Phys. Rev. B} {\bf V40-I6} (1989), 3616-3621.

 \bibitem{Shockley37} W. Shockley, ``The Empty Lattice Test of the Cellular Method in Solids'',
 {\it Phys. Rev. }{\bf 52} (1937), 866-872.

\bibitem{Janak71} {\it Computational methods in band theory}, edited by
P. M. Marcus, J. F. Janak, A. R. Williams (Plenum, New York, 1971), 323-339.

\bibitem{Gilat81} G. Gilat,
``General analytic method of zone integration for joint densities of states in metals'',
{\it Phys. Rev. B} {\bf V26-N4} (1981), 2243-2246.

\bibitem{Gilat66} G. Gilat \& L.J. Raubenheimer
``Acurate Numerical Method for Calculating Frequency distribution Functions in Solids'',
{\it Phys. Rev.}{\bf V144, N2} (1966), 390-395 ; erratum: G. Gilat \& L.J. Raubenheimer
``Acurate Numerical Method for Calculating Frequency distribution Functions in Solids'',
{\it Phys. Rev.}{\bf V147, N2} (1966), 670.

\bibitem{Feynman39} R. Feynman,
``Forces in molecules'',
{\it Phys. Rev.} {\bf V56-I4} (1939) 340-343.

\bibitem{Lehmann7072} G. Lehmann, P. Rennert, M. taut, H. Wonn,
{\it Phys. Status Solidi } {\bf 37} (1970) K27 ; G. Lehmann, M. Taut,
{\it Phys. Status Solidi } {\bf 54} (1972) 469.

\bibitem{Jepsen71} O. Jepsen and O. K. Andersen,
{\it Solid State Comm. } {\bf 9} (1971), 1763.

\bibitem{Gilat72} G. Gilat,
``Interpolation versus extrapolation in the Brillouin zone integration schemes'',
{\it Phys. Rev. B} {\bf V7-N2} (1972), 891-892.

\bibitem{Gilat75} G. Gilat, N. R. Bharatiya,
``Tetrahedron method of zone integration: inclusion of matrix elements'',
{\it Phys. Rev. B}{ \bf V12-N8} (1975), 3479-3481.

\bibitem{Gilat72b} G. Gilat,
``Analysis of methods for calculating spectral properties in solids'',
{\it J. Comp. Phys. }{ \bf 10} (1972), 432-465.

\bibitem{Morita71} T. Morita, T. Horiguchi,
``Lattice Green's functions for the cubic lattices in terms of the complete elliptic integral'',
{\it J. Math. Phys. } {\bf V12-N6} (1971), 981-986.

\bibitem{Oyamatsu93}  K. Oyamatsu,
``Nuclear shapes in the inner crust of a neutron stars'',
{\it Nucl. Phys. A} {\bf 561} (1993)  431-452.

 \bibitem{Koster57} G. F. Koster, ``Space Groups and Their Representations'',
{Solid State Physics}, {\bf Vol 5}, F. Seitz \& D. Turnbull editors,
Academic Press Inc., Publishers New York (1957).

\bibitem{Dutta2004} A. K. Dutta, M. Onsi and J.M. Pearson,
``Proton shell effects in neutron star matter'',
{\it Phys. Rev. C }{\bf 69} (2004), 052801.

\bibitem{Magier2003} P. Magierski, A. Bulgac, P.-H. Heenen,
``Exotic nuclear phases in the inner crust of neutron stars in the light
of the Skyrme-Hartree-Fock theory'',
{\it Nucl. Phys.} {\bf A719} (2003) 217-220.

\end{thebibliography}
\end{document}